\documentclass[twocolumn,showpacs, amsmath,amssymb]{revtex4}
\usepackage{graphicx}
\usepackage{dcolumn}
\usepackage{bm}

\begin{document}

\title{Crossover of the weighted mean fragment mass scaling in 2D brittle fragmentation}
\author{Hiroaki Katsuragi}
\email{katsurag@asem.kyushu-u.ac.jp}
\author{Daisuke Sugino}
\author{Haruo Honjo}
\affiliation{Department of Applied Science for Electronics and Materials, Interdisciplinary Graduate School of Engineering Sciences, Kyushu University, 6-1 Kasugakoen, Kasuga, Fukuoka 816-8580, Japan}
\date{\today}

\begin{abstract}
We performed vertical and horizontal sandwich $2D$ brittle fragmentation experiments. The weighted mean fragment mass was scaled using the multiplicity $\mu$. The scaling exponent crossed over at $\log \mu_c \simeq -1.4$. In the small $\mu (\ll\mu_c)$ regime, the binomial multiplicative (BM) model was suitable and the fragment mass distribution obeyed log-normal form. However, in the large $\mu (\gg\mu_c)$ regime, in which a clear power-law cumulative fragment mass distribution was observed, it was impossible to describe the scaling exponent using the BM model. We also found that the scaling exponent of the cumulative fragment mass distribution depended on the manner of impact (loading conditions): it was $0.5$ in the vertical sandwich experiment, and approximately $1.0$ in the horizontal sandwich experiment.

\end{abstract}


\maketitle

The origin of the power-law distribution in brittle fragmentation is one of the best-examined problems in statistical physics \cite{Beysens1,Turcotte1}. It has been examined in many recent experiments and simulations \cite{Ishii1,Oddershede1,Meibom1,Kadono1,Kadono2,Hayakawa1,Astrom4,Astrom1,Astrom2,Astrom3,Kun1,Kun2,Kun3,Diehl1}. In particular, the universality of fragmentation transition and low-impact energy fragmentation have been discussed \cite{Astrom1,Kun1}. Due to the success of scaling theory with critical phenomena, it is natural to consider the universality of critical behavior for various phenomena. Kun and Herrmann discussed the possibility of percolation universality using a point impacted granular solid model \cite{Kun1}. They also investigated the universality of shell fragmentation \cite{Kun2}. {\AA}str\"om et al.\ proposed another universality law for LJ liquid and elastic beam models \cite{Astrom1}. Dimensional analyses of the exponent of the power-law distribution have also been derived \cite{Hayakawa1,Kadono1,Astrom2,Astrom3}.

Previously, we conducted $2D$ brittle fragmentation experiments in which we applied a flat impact to one side of the specimen \cite{Katsuragi1}. This consisted of a vertical sandwich procedure using glass tubes. We showed that the critical scaling differed from that of percolation transition, and proposed a binomial multiplicative (or biased cascade) model for critical fragmentation. The binomial multiplicative (BM) model is very similar to the turbulent multifractal $p$ model \cite{Meneveau1}. This implies the similarity between brittle fragmentation and turbulence by means of multifractality. However, the BM model included a fitting parameter that was fixed at $a=2/3$, although the origin of this value was not clear. When a more realistic case was considered, the model predictions did not fit the experimental results \cite{Katsuragi2}. The model results also did not follow the power-law fragment mass distribution; rather, they obeyed a log-normal distribution due to the central limit theorem.

Low-impact energy fragmentation measured in experiments that involved dropping a $1D$ glass rod yielded log-normal distributions in the relatively low-impact energy regime \cite{Ishii1}. The log-normal form has also been observed in the $3D$ numerical results of viscoelastic crystal fragmentation \cite{Hayakawa1}. The first discussion of the log-normal distribution for a fragmentation process is found in Kolmogoloff \cite{Kolmogolov1}. It is not clear how the fragment mass distribution approaches the power-law form from the log-normal distribution. Do the fragments obey any other distributions before they reach the power-law form? The relation between the universal scaling law, the log-normal model, and multifractality is one of the most frequently discussed topics, even in the turbulent energy cascade problem \cite{Fisch1}. Since the brittle fragmentation phenomenon is very simple, it is very useful to investigate the origin of and path to the power-law form.

In order to study this problem, we performed low-impact energy fragmentation experiments. In addition to the glass tube results we reported previously \cite{Katsuragi1,Katsuragi2}, we also analyze the results for glass plate samples, which correspond to a horizontal sandwich procedure.

The experimental apparatus was very simple. Samples were sandwiched between a stainless steel plate and a stainless steel stage. Then, a heavy brass weight was dropped along guide poles. This experimental system was described in Ref.\ \cite{Katsuragi1}. After fragmentation, all the fragments were collected and their masses were measured using an electronic balance. We broke 25 new glass plates. Fifteen were $30$ mm $\times$ $30$ mm $\times$ $0.1$ mm in size, and ten were $60$ mm $\times$ $60$ mm $\times$ $0.1$ mm in size. We set the measurement limit for the minimum mass at $0.001$ g, but only analyzed the data for fragments down to $m_{\min}=0.01$ g. This $m_{\min}$ value is same as that used in the glass tube experiments. The glass tubes and plates corresponded to vertical and horizontal sandwich procedures, respectively.

\begin{figure}
\scalebox{0.9}[0.9]{\includegraphics{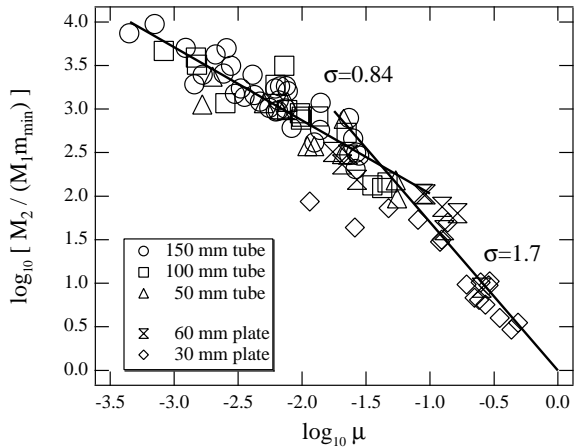}}
\caption{The log of the dimensionless weighted mean fragment mass $[M_2/(M_1 m_{\min})]$ as a function of the log of the pseudo-control-parameter $\mu$. The scaling crossed over from a log-normal to a power-law distribution regime around $\log \mu_c \simeq -1.4$. While $\sigma$ satisfied $\sigma\leq 1$ in the small $\mu$($\ll\mu_c$) regime, it exceeded $1$ in the $\mu \gg \mu_c$ regime. This implies that the BM model is unsuitable for the large $\mu$ regime.}
\label{fig:allM2M1}
\end{figure}

Let us introduce a critical divergence of the weighted mean fragment mass,
\begin{equation}
\frac{M_2}{M_1 m_{\min}} \sim \mu^{-\sigma},
\label{eq:M2M1def}
\end{equation}
where $M_k$ and $\mu$ are written as
\begin{equation}
M_k = \sum_{m} m^{k}n(m), \;\;\;\;\;\;\;\;\; \mu=m_{\min} \frac{M_0}{M_1}.
\label{eq:Mkmudef}
\end{equation}
where $m$ and $n(m)$ denote the fragment mass and fragment number of the mass $m$, respectively. Note that the summation in Eq.\ (\ref{eq:Mkmudef}) includes the largest fragment mass. The left-hand side of Eq.\ (\ref{eq:M2M1def}) also includes the factor $m_{\min}^{-1}$, which was not considered in the previous definition of $\sigma$ (Eq.\ (4) in Ref.\ \cite{Katsuragi1}). This factor is a normalization term for the weighted mean fragment mass and gives a dimensionless value. It does not affect the value of the scaling exponent. The multiplicity parameter $\mu$ was first introduced by Campi as a pseudo-control-parameter to analyze nuclear fragmentation \cite{Campi1}. It indicates the dimensionless normalized fragment number.

\begin{figure}
\scalebox{0.9}[0.9]{\includegraphics{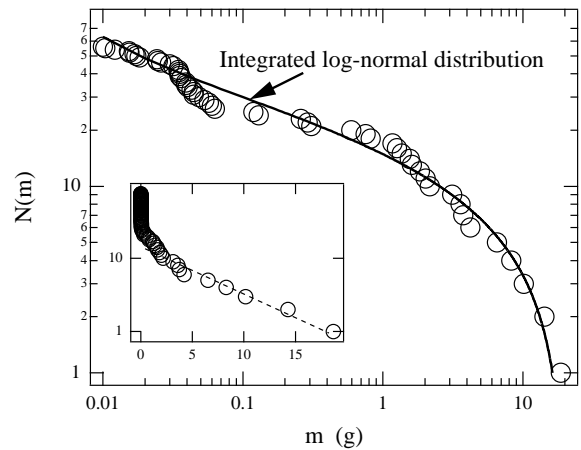}}
\caption{Log-normal form of the fragment mass distribution in the low-impact energy regime ($\log \mu = -2.55$ data). The sample was $150$-mm long glass tube. The inset shows a semi-log plot of the same distribution. The dashed line indicates an exponential-like tail.}
\label{fig:log-normal}
\end{figure}

The entire plot of $\log [M_2/(M_1 m_{\min})]$ vs. $\log \mu$ is shown in Fig.\ \ref{fig:allM2M1}. The figure shows that the scaling crosses over around ($\log \mu_c \simeq -1.4$).
There are also two divergent points in Fig.\ \ref{fig:allM2M1}, which are likely due to experimental failure, such as an oblique impact. However, we did not remove these points, since we do not have clear criteria to distinguish between success and failure. 
In the regime $\mu \ll \mu_c$, the scaling exponent $\sigma$ can be described using the previously obtained value $\sigma = 0.84 \leq 1$ \cite{Katsuragi1}. The higher-order weighted mean fragment mass exhibited a multi-scaling nature and its exponent agreed with the one predicted by the BM model. Therefore, we expect the fragment mass distribution to obey the log-normal form in this regime. Figure \ref{fig:log-normal} shows an integrated log-normal form of the cumulative fragment mass distribution $N(m)=\int_{m}^{\infty}n(m')dm'$ for a typical low-impact energy fragmentation ($150$-mm long glass tube data with $\log \mu = -2.55$). The integrated log-normal function can be written as
\begin{equation}
N(m)= A \int_{m}^{m_{\infty}} \frac{ \exp \left[ - \{\log (m'/\bar{m})\}^2 / 2\sigma_{ln}^2 \right]}{m' \sqrt{2\pi\sigma_{ln}^2} }dm'
\end{equation}
where $A$, $\bar{m}$, and $\sigma_{ln}$ are parameters, which were taken as $0.24$, $10.0$, and $2.0$ for the solid line in Fig.\ \ref{fig:log-normal}, respectively. We used $m_{\infty}=20$ as the cutoff scale. Since good agreement was obtained, the fragment mass distribution in $\mu \ll \mu_c$ followed a log-normal distribution.

\begin{figure}
\scalebox{0.9}[0.9]{\includegraphics{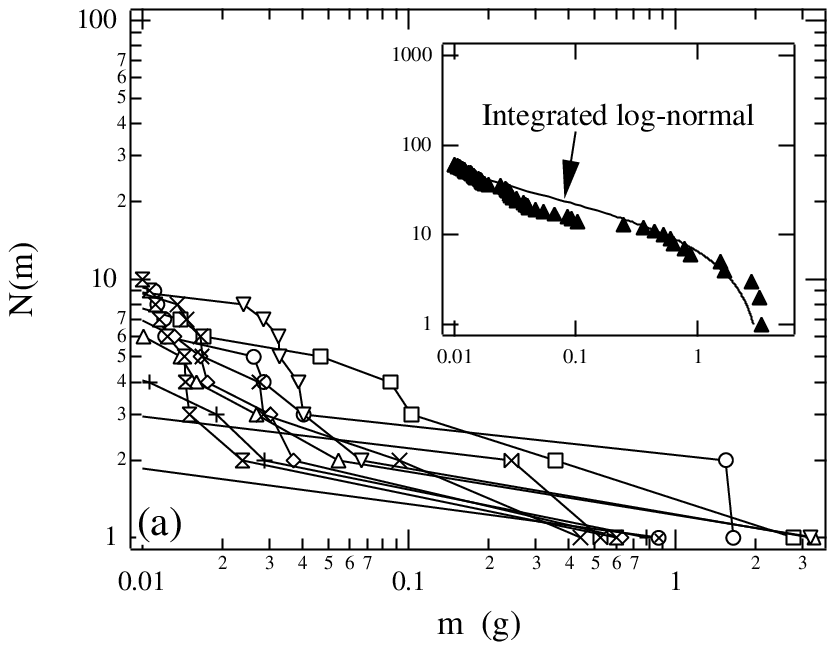}}
\scalebox{0.9}[0.9]{\includegraphics{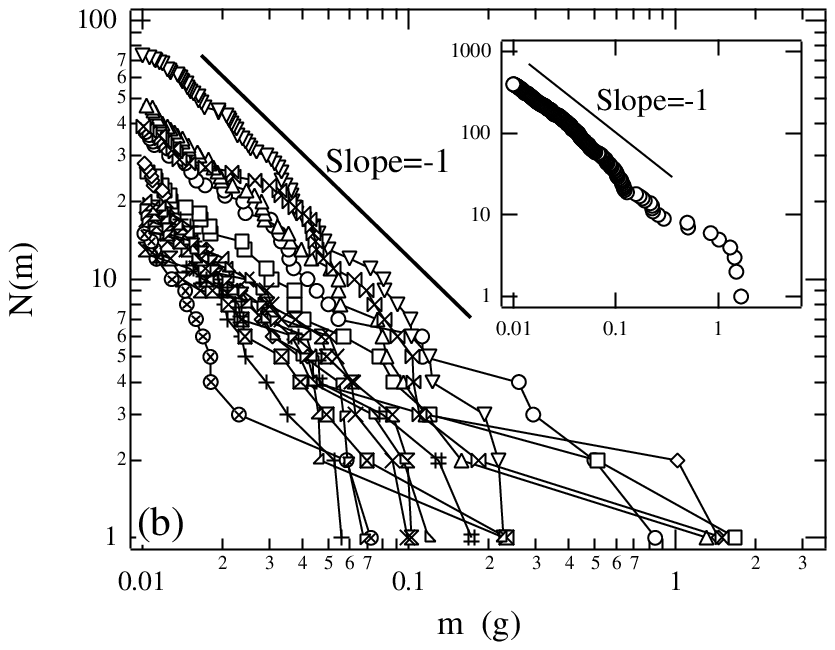}}
\scalebox{0.9}[0.9]{\includegraphics{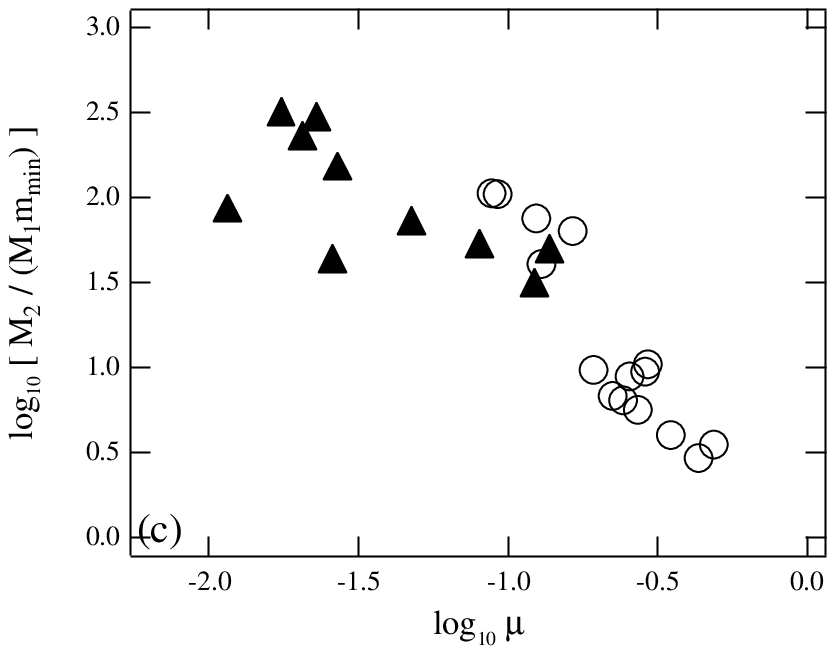}}
\caption{The cumulative fragment mass distribution of the glass plate samples in (a) the low-impact energy regime and (b) the high-energy-impact energy regime. (c) Scaling plot of $\log [M_2/(M_1 m_{\min})]$ vs. $\log \mu$ for all glass plate samples. The triangles correspond to the low-impact-energy cases (a), and the circles correspond to the high-impact-energy cases (b). The insets of (a) and (b) show the all summed curves.}
\label{fig:platecums}
\end{figure}

While most of the data exhibited a log-normal form, there were a small number of fragments in the low-impact energy regime in general (e.g., raw curves in Fig.\ \ref{fig:platecums}(a)) so that it was difficult to establish the form of the distribution directly. Therefore, we measured the weighted mean fragment mass using the moment $M_k$ of the distribution to obtain sufficient evidence.
However, we encountered problems when calculating the multiscaling exponent $\sigma_k$ (defined as $M_{k+1}/(M_k m_{\min}) \sim \mu^{-\sigma_k}$) for the glass plate data due to the large fluctuations in $M_{k+1}/(M_k m_{\min})$. We did not obtain reliable estimates of $\sigma_k$ for the glass plates, particularly in the large $k$ region. Therefore, we focused only on $M_2/(M_1 m_{\min})$ scaling here. The scaling in the large $k$ regime is obviously determined mainly by the largest fragment. This means that we used mean mass statistics instead of the largest mass statistics, in thia paper.

For the glass tubes, some fragmentation results showed a power-law distribution in the relatively large $\mu$ regime \cite{Katsuragi1}. In such regime, $\mu$ was close to $\mu_c$, i.e., the crossover might have already occurred (Fig.\ 4(b) in Ref.\ \cite{Katsuragi1}). Due to the dimensional restrictions of the experimental apparatus, we could only examine the small $\mu$ regime for glass tubes. Therefore, the clear crossover found in the glass plate fragmentation data has not been observed previously.

Another important characteristic is the power-law form of the cumulative fragment mass distribution for a fully fragmented state. It had different exponents for the tube and plate experiments. Figure \ref{fig:platecums} shows the cumulative fragment mass distributions in the range $m\geq 0.01$ g for the glass plate samples. Figures \ref{fig:platecums}(a) and \ref{fig:platecums}(b) give the low- and high-impact energy regime distributions, respectively. Each curve represents different imparted energy (dropping height of the weight) state.  The cumulative distributions of well-fragmented events (Fig.\ \ref{fig:platecums}(b)) have a power-law portion $N(m) \sim m^{-(\tau -1)}$ with an exponent $\tau-1$ of about $1$. Some distributions in Fig.\ \ref{fig:platecums}(b) contain large fragments, so that the scaling regions are restricted to almost one order of magnitude; however, most portions of the distributions follow $\tau-1=1$. The glass tube experiments had $\tau-1=0.5$ \cite{Katsuragi1}. This difference between the tubes and plates indicates that the exponent $\tau$ depends on the fracturing method. The value $\tau-1=1$ does not concur with the value predicted by Hayakawa and {\AA}str\"om et al., $\tau -1(=(d-1)/d )=1/2$ (for $d=2$) \cite{Astrom1,Hayakawa1}. They considered the propagating and branching dynamics of the crack (or the failure wave). Therefore, in the horizontal sandwich fragmentation of glass plates, mechanisms other than crack dynamics might determine the value of $\tau$.
Moreover, the boundary conditions are different between our horizontal sandwich experiments and the simulations of {\AA}str\"om et al. 
By contrast, Behera et al.\ obtained a value of $\tau-1 \simeq 1$ in the highly fragmented state for a lateral impact disk fragmentation simulation \cite{Kun3}. This value agrees with our experiments, despite the difference in the loading conditions. Kadono discussed the energy balance and obtained the inequality $1/2<\tau-1<1$ \cite{Kadono1}. This inequality range is also close to our result. 

On the other hand, our distributions in the low-impact energy regime showed the remains of large fragments and were rather flat (Fig.\ \ref{fig:platecums}(a)). This behavior resembles that of the integrated log-normal form, as described in Fig.\ \ref{fig:log-normal}. Although all the curves in Figs.\ \ref{fig:platecums}(a) and \ref{fig:platecums}(b) correspond to different imparted energy states, we tried summing up those. As a result, summed curves are shown in the insets; these more clearly indicate the integrated log-normal and power-law distributions. The solid curve in Fig.\ \ref{fig:platecums}(a) is same as that in Fig.\ \ref{fig:log-normal}, except for the cutoff scale $m_{\infty}=3.5$.

The weighted mean fragment mass scaling of the glass plate samples only are shown in Fig.\ \ref{fig:platecums}(c). Here, the triangles correspond to the low-impact energy regime distributions (Fig.\ \ref{fig:platecums}(a)), and the circles correspond to the high-impact energy regime distributions (Fig.\ \ref{fig:platecums}(b)). As expected, we confirmed a distinct separation between the two regimes using the weighted mean fragment mass and the multiplicity. The value of $\sigma$ became large ($\sigma \simeq 1.7$) in the larger $\mu$($\gg\mu_c$) regime (Fig.\ \ref{fig:allM2M1}). Although the BM model can be applied to small $\sigma$($\leq 1$) values, it is inappropriate for large $\sigma$($>1$). Furthermore, other models, such as the distributed and remaining cascade model, also break down for large $\sigma$ values \cite{Katsuragi2}. The point $\mu_c$ indicates the distribution crossover from the log-normal to the power-law. We cannot explain what happens in the large $\mu$ regime at present. Perhaps the smallest limit of the splitting mass might appear above $\mu_c$, similar to an idea proposed by Matsushita and Sumida \cite{Matsushita1}.
We assumed that the crossover point was universal, but there appears to be a slight difference between the points shown in Figs.\ \ref{fig:allM2M1} and \ref{fig:platecums}(c).
More details and direct observations of fragmentation are necessary to understand the crossover precisely. Theoretical studies are also required. In particular, an analysis in the vicinity of $\mu_c$ would be interesting to see how the transition occurs.

The crossover in Fig.\ \ref{fig:allM2M1} is reasonable from the viewpoint of the limit point. The limit point ($\log \mu$, $\log [M_2/(M_1m_{\min})]$)$=$($0$, $0$) corresponds to the completely fragmented state. In such a state, all fragments are the smallest unit size fragments. While it is extremely difficult to achieve such a state (i.e., fragment mass distributions exhibit power-laws in general), it can exist as an ideal limit case. If the BM scaling stretches until $\log \mu=0$ in Fig.\ \ref{fig:allM2M1}, $\log [M_2/(M_1m_{\min})]$ never reaches the value $0$. This is a nonphysical state. Therefore, it is natural that the crossover point corresponds to a certain value $\mu_c$.

{\AA}str\"om et al.\ recently proposed a generic fragment mass distribution form that was composed of a power-law portion and an exponential portion \cite{Astrom2}. The former originates from the dynamics of crack branching and merging, and the latter results from the Poisson process. Their proposed form also applies to the low-impact energy regime. The inset in Fig.\ \ref{fig:log-normal} depicts a semi-log plot of the same $N(m)$ distribution that was explained using a log-normal distribution. It shows a straight (i.e., exponential) tail, which suggests that the {\AA}str\"om model may also be suitable. However, from the viewpoint of the multiscaling nature of critical fragmentation, the BM model and log-normal distribution are more plausible.
Diehl et al.\ obtained a similar coincidence between $2D$ explosive fragmentation simulations and the BM model \cite{Diehl1}. They also discussed the log-normal distribution form. 

Very recently, Wittel et al.\ reported the results of shell fragmentation experiments and simulations \cite{Kun2}. They concluded that the {\it impact} fragmentation of shells showed a continuous transition, while the {\it explosive} one showed an abrupt transition. In our experiments, fragmentation seemed to occur suddenly. We could not obtain samples that only had visible macro-cracks, but did not split. A small amount of imparted energy cannot make brittle solids cleave. This might imply a ``latent-heat-like behavior''. That is, the beginning of fragmentation requires a finite ``latent energy'' to generate macro-cracks. The splitting occurs abruptly and it proceeds according to the BM model statistics. We can observe critical scaling in the range $\mu > 0$. However, we cannot discuss the scaling in the range $\mu < 0$, since it corresponds to the unfragmented state. The fragmentation transition of open $2D$ objects involved in flat impacts is not yet understood very well in terms of the phase transition, and this is still an open question. Conversely, the transition from the log-normal to the power-law is characterized by the crossover of the weighted mean fragment mass scaling, as demonstrated above.

Wittel et al.\ also revealed that the scaling exponent $\tau$ is dependent on the loading conditions in numerical simulations \cite{Kun2}. While this was not consistent with their experimental results, it concurs with our findings qualitatively if we consider the vertical and horizontal sandwich procedures to correspond to impact and explosive fragmentation processes, respectively. Quantitatively, their values of $\tau$ differed from ours slightly. This might result from the difference between an open $2D$ sample and a closed shell sample.

In summary, we examined $2D$ brittle fragmentation using experiments with glass tubes and glass plates. The exponent $\tau$ had different values depending on the loading conditions, which consisted of either a horizontal or vertical sandwich impact to the $2D$ surface. Contrarily, the normalized weighted mean fragment mass scaling was universal and had a crossover point at which the fragment mass distribution changed from a log-normal to a power-law type. The results were consistent with other recent experiments and numerical simulations, but included new experimental findings about the relatively large $\mu$ weighted mean fragment mass scaling.

We thank Dr.\ J. A. {\AA}str\"om and Prof.\ H. Nakanishi for their helpful comments.
This research was partially supported by the Ministry of Education, Culture, Sports, Science and Technology, through a Grant-in-Aid for Young Scientists No.\ 16740206.

\end{document}